\documentclass{article}
\usepackage{amsmath,amssymb,amsthm,amssymb,graphicx}
 \newtheorem{thm}{Theorem}[section]

 \newtheorem{prop}[thm]{Proposition}
 \theoremstyle{definition}
 
 \theoremstyle{remark}

 \numberwithin{equation}{section}

\newcommand{\paae}{p^{a}_{\alpha_1}}
\newcommand{\paat}{p^{a}_{\alpha_2}}
\newcommand{\pabe}{p^{b}_{\beta_1}}
\newcommand{\pabt}{p^{b}_{\beta_2}}

\newcommand{\pbae}{p^{a}_{\alpha_1}}
\newcommand{\pbat}{p^{a}_{\alpha_2}}
\newcommand{\pbbe}{p^{b}_{\beta_1}}

\newcommand{\pbbeae}{p^{b|a}_{\beta_1\alpha_1}}
\newcommand{\pbbeat}{p^{b|a}_{\beta_1\alpha_2}}
\newcommand{\pbbtae}{p^{b|a}_{\beta_2\alpha_1}}
\newcommand{\pbbtat}{p^{b|a}_{\beta_2\alpha_2}}

\newcommand{\pabeat}{p^{a|b}_{\beta_1\alpha_2}}

\newcommand{\pbaebe}{p^{b|a}_{\alpha_1\beta_1}}

\newcommand{\paaebe}{p^{a|b}_{\alpha_1\beta_1}}
\newcommand{\paaebt}{p^{a|b}_{\alpha_1\beta_2}}
\newcommand{\paatbe}{p^{a|b}_{\alpha_2\beta_1}}
\newcommand{\paatbt}{p^{a|b}_{\alpha_2\beta_2}}
\newcommand{\sgn}[1]{\epsilon_{#1}}

\begin{document}

\title{On the consistency of the quantum-like representation algorithm for hyperbolic interference}

%\classification{ 03.65.Ca, 02.50.Cw}

\author{Peter Nyman\\ School of Computer Science, Physics and Mathematics \\ Linnaeus University, S-35195, Sweden\\ peter.nyman@lnu.se}
\maketitle

\begin{abstract}
Recently quantum-like representation algorithm (QLRA) was introduced
by A. Khrennikov \cite{K1}--\cite{K5} to solve the so-called
``inverse Born's rule problem'': to construct a representation of
probabilistic data by a complex or more general (in particular,
hyperbolic) probability amplitude which matches Born's rule or its
generalizations. The outcome from QLRA is coupled to the formula of
total probability with an additional term corresponding to
trigonometric, hyperbolic or hyper-trigonometric interference.  The
consistency of QLRA for probabilistic data corresponding to
trigonometric interference was recently proved \cite{P1}. We now
complete the proof of the consistency of QLRA to cover hyperbolic
interference as well. We will also discuss hyper trigonometric
interference. The problem of consistency of QLRA arises, because
formally the output of QLRA depends on the order of conditioning.
For two observables (e.g., physical or biological) $a$ and $b,$
$b\vert a$- and $a \vert b$- conditional probabilities produce two
representations, say in  Hilbert spaces  $H^{b\vert a}$ and
$H^{a\vert b}$  (in this paper over the hyperbolic algebra). We
prove that under ``natural assumptions'' these two representations
are unitary  equivalent (in the sense of hyperbolic Hilbert space).
\end{abstract}
%\keywords{Born's rule problem, hyperbolic interference, hyper trigonometric interference, inverse  order of conditioning, quantum-like  representation algorithm}
%%%%%%%%%%%%%%%%%%%%%%%%%%%%%%%%%%%%%%%%%%%%
%% MAINMATTER
%%%%%%%%%%%%%%%%%%%%%%%%%%%%%%%%%%%%%%%%%%%%
\section{Introduction}
The interrelation between classical and quantum probabilities was
early studied by von Neumann, see \cite{VN} and was followed by
methods to generalize the probability theory to include quantum
probabilities by Gudder, see \cite{GD1}--\cite{GD3}.  For more
recent and wide-ranging studies, see Svozil \cite{SV1}, \cite{SV2},
Fine \cite{F}, Garola et al. \cite{GAR1}--\cite{GAR3}, Dvurecenskij
and Pulmanova \cite{DV}, Ballentine \cite{BL}, O.  N\'an\'asiov\'a
et al \cite{NAN}, \cite{NAN1},  Allahverdyan et al \cite{AL},
Khrennikov  \cite{K6,K7}. The basic rule of QM is  the Born's rule.
Therefore the study of its origin is very important for quantum
foundations.  In a series of papers  \cite{K1} --\cite{K5}
Khrennikov studied so called ``inverse Born's rule problem'':

\medskip

{\bf IBP} (inverse Born problem): {\it To construct a representation
of probabilistic data (of any origin)  by a complex probability
amplitude which matches Born's rule.}

\medskip

The solution of IBP provides a possibility to represent
probabilistic data by ``wave functions'' and operate with this data
by using linear algebra (as we do in conventional QM). However, as
it was found  in \cite{K1} --\cite{K5}, some data do not permit the
complex wave representation. In this case probabilistic amplitudes
valued in the hyperbolic algebra (a two dimensional Clifford
algebra) should be used as well. A special algorithm, quantum-like
representation algorithm (QLRA), was created to transfer
probabilities into probabilistic amplitudes. Depending on the data
these amplitudes are {\it complex, hyperbolic or hyper-complex.}

\medskip

Formally, the  output of  QLRA depends on the order of conditioning
of probabilities. For two observables $a$ and $b,$  $b\vert a$- and
$a \vert b$- conditional probabilities produce two representations,
say in  Hilbert spaces (complex or hyperbolic) $H^{b\vert a}$ and
$H^{a\vert b}$. In this paper we will be interested in hyperbolic
amplitudes as outputs of QLRA and therefore consider the hyperbolic
Hilbert space. The case of complex amplitudes has already been
studied in \cite{K1}. It was shown that under ``natural assumptions"
these two conditional probabilistic representations are unitary
equivalent. This result proved the consistency of QLRA for complex
amplitudes, now we will study the case of hyperbolic amplitudes.

In a purely mathematical  framework the problem of consistency of
two representations is nothing else than construction of a special
unitary operator in hyperbolic Hilbert space establishing the
equivalence of two representations. This paper is also a
contribution to mathematical physics over hyperbolic numbers, see,
e.g., \cite{Baylis}--\cite{Kisil1}.

\section{Inversion of Born's Rule}

We consider the simplest situation. There are given two
dichotomous observables of any context: $a=\alpha_1, \alpha_2$ and $b=\beta_1,
\beta_2.$  We set $X_a=\{\alpha_1,
\alpha_2\}$ and $X_b=\{\beta_1, \beta_2\}$ -- ``spectra of
observables".

We assume that there is given  the matrix of transition
probabilities ${\bf P}^{b\vert a}= (p^{b\vert a}_{\beta \alpha}),$ where
$p^{b\vert a}_{\beta \alpha}\equiv P(b=\beta \vert a=\alpha)$ is the
probability to obtain the result $b=\beta$ under the condition
that the result $a=\alpha$ has been obtained. There are also given probabilities
$p^a_\alpha \equiv P(a=\alpha),
\alpha\in X_a,$ and $p^b_\beta \equiv P(b=\beta), \beta \in X_b.$
Probabilistic  data $C=\{ p^a_{\alpha}, p^b_{\beta}\}$ are related
to some experimental context (in the physics preparation procedure).

IBP is to represent this data by a probability amplitude $\psi$  (in
the simplest case it is complex-valued, but we are interested in
more general amplitudes) such that Born's rule  holds for both
observables:
\begin{equation}
\label{BR} p^b_{\beta}= \vert \langle \psi, e^{b\vert a}_\beta\rangle
\vert^2\,,\qquad p^a_{\alpha}= \vert \langle \psi, e^{b\vert a}_\alpha\rangle
\vert^2\,,
\end{equation}
where $\{e^{b\vert a}_\beta \}_{\beta \in X_b}$ and $\{e^{b\vert
a}_\alpha\}_{\alpha \in X_a}$ are orthonormal bases for observables
$b$ and $a,$ respectively (so the observables are diagonal in the
respective bases).

In \cite{K1} --\cite{K5} the solution of IBP  was given in the form
of an algorithm which constructs a probability amplitude from the
data. Formally, the output of this algorithm depends on the order of
conditioning. By starting with the matrix of transition
probabilities ${\bf P}^{a\vert b},$ instead of ${\bf P}^{b\vert a},$
we construct another probability amplitude $\psi^{a\vert b}$ (the
amplitude in (\ref{BR}) should be denoted by   $\psi^{b\vert a})$
and other bases, $\{e^{a\vert b}_\beta\}_{\beta \in X_b}$ and
$\{e^{a\vert b}_\alpha\}_{\alpha \in X_a}.$ We shall see that under
natural assumptions these two representations are unitary
equivalent.

\section{QLRA}
\subsection{$H^{b \vert a}$-conditioning}

Suppose that the matrix of transition probabilities ${\bf P}^{b\vert
a}$ is given.  In \cite{K1} --\cite{K5} the following formula for
the interference of probabilities (generalizing the classical
formula of total probability)
 was derived:
\begin{equation} 
\label{PBbeta} p^b_{\beta} \;=\; \sum_\alpha p^a_{\alpha} p^{b\vert a}_{\beta
\alpha} + 2 \lambda_\beta \; \sqrt{\prod_\alpha p^a_{\alpha}
p^{b\vert a}_{\beta \alpha}},
\end{equation}
where the ``coefficient of interference" is given by
\begin{equation}
\label{GGTT}
\lambda_\beta \;=\; \frac{p^b_{\beta} - \sum_\alpha
p^a_{\alpha} p^{b\vert a}_{\beta \alpha}}{2 \sqrt{\prod_\alpha p^a_{\alpha}
p^{b\vert a}_{\beta \alpha}}}\, .
\end{equation}
We will  proceed under the conditions:

(1) ${\bf P}^{b\vert a}$ is doubly stochastic (for not doubly stochastic matrix, see section \ref {VBR}).

(2) Probabilistic data $C=\{ p^a_{\alpha}, p^b_{\beta}\}$ consist
of strictly positive probabilities.

(3) The absolute values of the coefficients of interference
$\lambda_\beta, \beta \in X_b,$ are larger than one: $ \vert
\lambda_\beta\vert > 1\,. $

Probabilistic data $C$ such  that $\vert \lambda_\beta\vert \leq 1$
are called {\it trigonometric}. In this case we have the
conventional formula of trigonometric interference: $$ p^b_{\beta}
\;=\; \sum_\alpha p^a_{\alpha} p^{b\vert a}_{\beta \alpha} + 2
\lambda_\beta \; \sqrt{\prod_\alpha p^a_{\alpha} p^{b\vert a}_{\beta
\alpha}}\,, $$ where
\begin{equation}
\label{GGTT1}
\lambda_\beta=\cos \theta_\beta\,.
\end{equation}
The case of trigonometric interference  (i.e. $\vert
\lambda_\beta\vert \leq 1$) has been studied in \cite{P1}.
Therefore, now we consider the case of hyperbolic interference:
$\vert \lambda_\beta\vert >1.$  We represent this coefficient of
interference by
\begin{equation}\label{hyplambda}
\lambda_\beta=\epsilon_{\beta}\cosh \theta_\beta
\end{equation}
 where $\epsilon_{\beta}=\text{sign } \lambda_{\beta}$.

Furthermore, in the case of hyper-trigonometric interference (i.e.
$\vert \lambda_{\beta_i}\vert >1$ and $\vert \lambda_{\beta_j}\vert
\leq 1$) we have $\lambda_{\beta_i}=\cos \theta_{\beta_i}$ and
$\lambda_{\beta_j}=\epsilon_{\beta_j}\cosh \theta_{\beta_j} \, ,$
where $\epsilon_{\beta_j}=\text{sign } \lambda_{\beta_j}
,\;i,j\in\{1,2\},i\neq j$.
\medskip
\begin{prop}
Let ${\bf P}^{b\vert a}$ be doubly stochastic. Then the case of
mixed hyper-trigonometric interference is excluded.
 \end{prop}

This proposition follows straightforward from the equality
$$\lambda_{\beta_1}+\lambda_{\beta_2}=0$$  and  the condition
that ${\bf P}^{b\vert a}$ is doubly stochastic, see \eqref{GGTT}. We
have:
\begin{eqnarray}
 \label{GGTT2}
 \lambda_{\beta_1}+\lambda_{\beta_2} &=& \frac{p^b_{\beta_1} - \sum_\alpha
 p^a_{\alpha} p^{b\vert a}_{\beta_1 \alpha}}{2 \sqrt{\prod_\alpha p^a_{\alpha}
 p^{b\vert a}_{\beta_1 \alpha}}}+\frac{p^b_{\beta_2} - \sum_\alpha
 p^a_{\alpha} p^{b\vert a}_{\beta_2 \alpha}}{2 \sqrt{\prod_\alpha p^a_{\alpha}
 p^{b\vert a}_{\beta_2 \alpha}}}
\\\nonumber
  &=& \frac{p^b_{\beta_1} +p^b_{\beta_2}- p^a_{\alpha_1}\sum_\alpha
  p^{b\vert a}_{\beta_1 \alpha} - p^a_{\alpha_2}\sum_\alpha
  p^{b\vert a}_{\beta_1 \alpha}}{2 \sqrt{\prod_\alpha p^a_{\alpha}
 p^{b\vert a}_{\beta_1 \alpha}}}
 \\\nonumber
&=&0\, .
 \end{eqnarray}
There is a contradiction between \eqref{GGTT2} and the definition of
hyper-trigonometric interference:$\vert \lambda_{\beta_i}\vert >1$
and $\vert \lambda_{\beta_j}\vert \leq 1$.

\bigskip

Therefore, we will focus on hyperbolic interference (since
trigonometric interference  has already been studied)   and
introduce the {\it hyperbolic algebra} ${\bf G}$; see appendix and
\cite{K4a}. Denote its generator (different from unit 1) by  $j:$
\[j^2=1.\]
An element of ${\bf G}$ can be represented as  $z=x+j y, \, x,y\in
\mathbb{R}.$
 We introduce the hyperbolic exponential function
\begin{equation}\label{HEF}
e^{j \theta}=\cosh \theta+j \sinh\theta, \quad \theta \in \mathbb{R}.
\end{equation}
Define also $\bar {z}=x-j y$, it is apparent that $\bar {z}\in{\bf G}$.
We also use the identities
\begin{equation}\label{EF2}
\cosh\theta=\frac{e^{j \theta}+e^{-j \theta}}{2},\quad \sinh\theta=\frac{e^{j \theta}-e^{-j \theta}}{2j}.
\end{equation}
Thus, by using the elementary formula:
\begin{equation}\label{HEF2} D=A+B\pm 2\sqrt{AB}\cosh \theta=\vert \sqrt{A}\pm e^{j
\theta}\sqrt{B}|^2,\quad A, B > 0\, , \theta\in \mathbb{R}\, ,
j^2=1\end{equation} for real numbers of $A$ and $B$, we can
represent the probability $p^b_{\beta}$ as the square of the
hyperbolic amplitude (Born's rule): $ p^b_{\beta}=\vert \psi^{b\vert
a}_\beta \vert^2 \,.$ Here
\begin{equation}
\label{EX1} \psi^{b\vert a}_\beta \;=\; \sqrt{p^a_{\alpha_1}p^{b\vert a}_{\beta \alpha_1}}
\pm  e^{j \theta_\beta} \sqrt{p^a_{\alpha_2} p^{b\vert a}_{\beta \alpha_2}}\,, \quad
\beta \in X_b\,.
\end{equation}
The formula (\ref{EX1}) gives the hyperbolic amplitude, the output
of QLRA for any probabilistic data $C$ if $|\lambda|>1$. This is the
normalized vector in the two dimensional hyperbolic Hilbert
space\footnote{For the definition of hyperbolic Hilbert space, see
appendix.}, say $H^{b\vert a}$:
\begin{equation}
\label{EXY}
\psi^{b \vert a}  =
\psi^{b\vert a}_{\beta_1} e^{b\vert a}_{\beta_1} + \psi^{b\vert a}_{\beta_2} e^{b\vert a}_{\beta_2},
\end{equation}
where
$
\label{BasTE} e^{b\vert a}_{\beta_1} \;=\;\left(  1 \;
0\right )^T,\quad e^{b\vert a}_{\beta_2} \;=\;\left(  0 \;
1\right )^T.$

To solve {\bf IBP} completely, we would like to have Born's rule not
only for the $b$-variable, but also for the $a$-variable:
$p^a_{\alpha}=\vert \langle \psi^{b\vert a}, e^{b \vert a}_\alpha
\rangle\vert^2 \;, \alpha \in  X_a.$ Here the $a$-basis in the
hyperbolic Hilbert space $H^{b\vert a}$  is given, see \cite{K1}
--\cite{K5} for details, by
\begin{equation}
 e^{b\vert a}_{\alpha_1} \;=\;\left( \begin{array}{l} \sqrt{p^{b\vert a}_{\beta_1 \alpha_1}} \\
\sqrt{p^{b\vert a}_{\beta_2 \alpha_1}}
\end{array}
\right ),\quad e^{b\vert a}_{\alpha_2} \;=\;\left( \begin{array}{l}  \; \sqrt{p^{b\vert a}_{\beta_1 \alpha_2}} \\
- \sqrt{p^{b\vert a}_{\beta_2 \alpha_2}}
\end{array}
\right )\,.
\end{equation}
This basis vectors are orthonormal, since ${\bf P}^{b \vert a}$ is
assumed to be doubly stochastic. In this basis the hyperbolic
amplitude $\psi^{b \vert a}$ is represented as
\begin{equation}
\label{PSI}
\psi^{b\vert a}= \sqrt{p^a_{\alpha_1}} e^{b\vert a}_{\alpha_1} \pm
e^{j \theta_{\beta_1}} \sqrt{p^a_{\alpha_2}} e^{b\vert a}_{\alpha_2}
\end{equation}
We recall that in QM two vectors (say $\psi'_1, \psi'_2$) define the
same state $\psi'$ if they differ by multipliers of the form
$c=e^{i\varphi}$\;(e.i. if $\psi'_1= e^{i\varphi}\psi'_2$ for some
$\varphi$ ). We will use a similar terminology for the case of the
hyperbolic algebra: two vectors $\psi_1, \psi_2$ define the same
state if $\psi_1=\pm e^{j\gamma}\psi_2$. The consistency of this
definition follows from the fact that $$|\psi_2|^2=|\pm
e^{j\gamma}|^2|\psi_2|^2=|e^{j\gamma}\psi_2|^2=|\psi_1|^2.$$ Thus
measurements on these two states produce the same probability
distribution.

Each hyperbolic amplitude $\psi^{b \vert a}$ produced by QLRA
determines a {\it quantum-like state} (representing given
probabilistic data) -- the equivalence class $\Psi^{b \vert a}$
being determined by the representative $\psi^{b \vert a}.$

\subsection{$H^{a \vert b}$-conditioning}

For ${a \vert b}$-conditioning the state is represented by
\begin{equation}
\label{EX1A} \psi^{a\vert b}_\alpha \;=\; \sqrt{p^b_{\beta_1}p^{a\vert b}_{\alpha \beta_1}}
\pm e^{j \theta_\alpha} \sqrt{p^b_{\beta_2} p^{a\vert b}_{\alpha \beta_2}}\,, \quad
\alpha \in X_a\,.
\end{equation}
 For any collection of probabilistic data $C$,
QLRA produces the hyperbolic  amplitude $ \psi^{a \vert b}$ if
$|\lambda|>1$ (the normalized vector in the two dimensional
hyperbolic Hilbert space, say $H^{a\vert b}):$
\begin{equation}
\label{EXYA}
\psi^{a\vert b}  =
\psi^{a\vert b}_{\alpha_1} e^{a\vert b}_{\alpha_1} +
\psi^{a\vert b}_{\alpha_2} e^{
a \vert b}_{\alpha_2},
\end{equation}
where $ \label{BasTET} e^{b\vert a}_{\beta_1} =\left(  1 \; 0\right
)^T,\quad e^{b\vert a}_{\beta_2} =\left(  0 \; 1\right )^T.$ Here
the $b$-basis in the hyperbolic  Hilbert space  $H^{a\vert b}$  is
given by
\begin{equation}
 e^{a\vert b}_{\beta_1} \;=\;\left( \begin{array}{l}
\sqrt{p^{a\vert b}_{\alpha_1 \beta_1}} \\
\sqrt{p^{a\vert b}_{\alpha_2 \beta_1}}
\end{array}
\right ),\, e^{a\vert b}_{\beta_2} \;=\;\left( \begin{array}{l}  \;
\sqrt{p^{a\vert b}_{\alpha_1 \beta_2}} \\
- \sqrt{p^{b\vert a}_{\alpha_2 \beta_2}}
\end{array}
\right )\,.
\end{equation}
In this basis the amplitude $\psi^{a\vert b}$ is represented as
\begin{equation}
\label{PSIA}
\psi^{a\vert b }= \sqrt{p^b_{\beta_1}} e^{a\vert b}_{\beta_1} \pm
e^{j \theta_{\alpha_1}} \sqrt{p^b_{\beta_2}} e^{b\vert a}_{\beta_2}
\end{equation}
As in the case of $H^{b \vert a}$-representation, the {\it
quantum-like state}  (representing given probabilistic data) is
defined as
 the equivalence class
 $\Psi^{a \vert  b}$  with the representative $\psi^{a \vert b}.$

\section{Unitary equivalence of $b \vert a$- and $a \vert b$-representa- tions}

Thus, as we have seen, by selecting two types of conditioning, we
represented the probabilistic data $C=\{p_\alpha^a, p_\beta^b\}$ by
two quantum-like states, $\Psi^{b \vert a}$ and $\Psi^{a \vert b}.$
We are interested in  the consistency of these representations.

We remark that any linear operator $W: H^{b \vert a} \to H^{a \vert
b}$ induces the map of equivalence classes of the hyperbolic unit
sphere\footnote{The hyperbolic  unit sphere is given by
$|e^{j\theta}|^2=\cosh^2{\theta}-\sinh^2{\theta}=1$} with respect to
multipliers  $c= \pm e^{j \gamma}.$ We define the unitary operator
$U_{b \vert a}^{a \vert b}: H^{b \vert a} \to H^{a \vert b}$ by
$U(e^{b \vert a}_\alpha)= e^{a \vert b}_\alpha, \alpha \in X_a.$ It
induces the mentioned map of equivalent classes.
\begin{thm}
The operator $U_{b \vert a}^{a \vert b}$ maps $\Psi^{b \vert a}$
into $\Psi^{a \vert b}$ if and only if the following interrelation
of symmetry takes place for the matrices of transition probabilities
${\bf P}^{b \vert a}$ and ${\bf P}^{a \vert b}$:
\begin{equation}
\label{PTTT}
p^{b \vert a}_{\beta \alpha}=p^{a \vert b}_{\alpha \beta },
\end{equation}
 for all $\alpha$ and $\beta$ from the spectra of observables $a$ and $b.$
\end{thm}

\begin{proof} Take the representative of $\Psi^{b \vert a}$ given by (\ref{PSI}).
Then
\begin{equation}\label{UPSI}
U_{b \vert a}^{a \vert b} \psi^{b \vert a}= \sqrt{p^a_{\alpha_1}} e^{a\vert b}_{\alpha_1} \pm
e^{j \theta_{\beta_1}} \sqrt{p^a_{\alpha_2}} e^{a \vert b}_{\alpha_2}
\end{equation}
Our aim is to show that this vector is
equivalent to the vector $\psi^{a \vert b}$ given by
(\ref{EXYA}).
The coefficients of interference  $\lambda_\alpha$ play in the $H^{a
\vert b}$-representation the same role as the coefficients of
interference  $\lambda_\beta$ played in $H^{b \vert
a}$-representation:
\begin{align}\label{PSI7}
\paae&={\pabe \paaebe}+{\pabt\paaebt}+
\sgn{\lambda_{\alpha_1}}2|\lambda_{\alpha_1}| \sqrt{\pabe \paaebe
\pabt\paaebt}\\\nonumber &\Leftrightarrow,\\\nonumber
\lambda_{\alpha_1}&=\frac{\paae-{\pabe \paaebe}-{\pabt\paaebt}}
{2\sqrt{\pabe \paaebe \pabt\paaebt}},
\end{align}
where  $\sgn{\lambda_{\alpha_1}}=\text{sign } \lambda_{\alpha_1}$.
We consider $|\lambda_{\alpha_1}|>1$ and thus
$|\lambda_{\alpha_1}|=\cosh{\theta_{\alpha_1}}$. We also calculate
\begin{eqnarray}\label{Psipsi}
\psi^{a|b}_{\alpha_2}\overline{\psi^{a|b}_{\alpha_1}}%&=\left(\sqrt{\pabe \paatbe}-e^{i \theta_{\alpha_1}}\sqrt{\pabt\paatbt}\right) \left(\sqrt{\pabe \paaebe}+e^{-i \theta_{\alpha_1}}\sqrt{\pabt\paaebt}\right)\\\nonumber
&=&\pabe\sqrt{\paaebe  \paatbe}\\\nonumber
&+&\sgn{\lambda_{\alpha_1}}\sgn{\lambda_{\alpha_2}}\pabt\sqrt{\paatbt\paaebt}\\\nonumber
&+&\sgn{\lambda_{\alpha_2}}(\cosh\theta_{\alpha_1}+j\sinh\theta_{\alpha_1})\sqrt{\pabt\paatbt\pabe \paaebe}\\\nonumber
&+&\sgn{\lambda_{\alpha_1}}(\cosh\theta_{\alpha_1}-j\sinh\theta_{\alpha_1})\sqrt{\pabe \paatbe\pabt\paaebt},
\end{eqnarray}
where $\psi^{a|b}_{\alpha_2}=\sqrt{\pabe \paatbe}\pm e^{j \theta_{\alpha_1}}\sqrt{\pabt\paatbt}$ is given by (\ref{PSIA}).
We also use  $$|
\psi^{a|b}_{\alpha_i}|^2=p^{a}_{\alpha_i}\Leftrightarrow
\psi^{a|b}_{\alpha_i}=\pm\sqrt{p^{a}_{\alpha_i}}
\left(\cosh\gamma_{\alpha_i}+j\sinh\gamma_{\alpha_i}\right),$$
where\footnote{see appendix for definition of the argument ($\arg$)
in the hyperbolic algebra.}
$$\gamma_{\alpha_i}=\arg{\psi^{a}_{\alpha_i}}, i\in\{1,2\}$$ and this
implies
\begin{equation}\label{cosgamma}
\psi^{a|b}_{\alpha_2}\overline{\psi^{a|b}_{\alpha_1}}%=\sqrt{\paae\paat}e^{i\left(\gamma_{\alpha_2}-\gamma_{\alpha_1}\right)}
=\pm\sqrt{\paae \paat}\left(\cosh\left(\gamma_{\alpha_2}-\gamma_{\alpha_1}\right)+j\sinh\left(\gamma_{\alpha_2}-\gamma_{\alpha_1}\right)\right).
\end{equation}
The real parts of the equations (\ref{Psipsi}) and (\ref{cosgamma})
give:
\begin{align}\label{ata}
\pm\sqrt{\paae \paat}\cosh\left(\gamma_{\alpha_2}-\gamma_{\alpha_1}\right)=\pabe\sqrt{\paaebe  \paatbe}-\pabt\sqrt{\paatbt\paaebt}
\\\nonumber
+\sgn{\lambda_{\alpha_1}}\cosh\theta_{\alpha_1}(\sqrt{\pabe \paatbe\pabt\paaebt}-\sqrt{\pabt\paatbt\pabe \paaebe}).
\end{align}
Notice that $\lambda_{\alpha_2}=-\lambda_{\alpha_1}$ in
\eqref{GGTT2} implies that
$$\sgn{\lambda_{\alpha_1}}=-\sgn{\lambda_{\alpha_2}},
\sgn{\lambda_{\alpha_1}}\sgn{\lambda_{\alpha_2}}=-1.
$$
Moreover, since $\pabt=1-\pabe$ and  ${\bf P}^{a\vert b}$ is doubly
stochastic, i.e.,
 $\paaebt=\paatbe=1-\paaebe=1-\paatbt$, we rewrite
(\ref{ata})
\begin{align}\label{nia}
\pm\sqrt{\paae \paat}\cosh\left(\gamma_{\alpha_2}-\gamma_{\alpha_1}\right)
%\\\nonumber&=\pabe\sqrt{\paaebe  (1-\paaebe)}-(1-\pabe)\sqrt{\paaebe(1-\paaebe)}
%&-\cos\theta_{\alpha_1}\left(\sqrt{(1-\pabe)\paaebe\pabe \paaebe}
%+\sqrt{\pabe (1-\paaebe)(1-\pabe)(1-\paaebe)}\right)\\\nonumber
&=\left(2\pabe-1\right)\sqrt{\paaebe(1-\paaebe)}\\\nonumber
&+\sgn{\lambda_{\alpha_1}}\cosh\theta_{\alpha_1}\left(1-2\paaebe\right)\sqrt{(1-\pabe)\pabe}.
\end{align}
Then by (\ref{GGTT}) and (\ref{hyplambda}) we obtain  $\cosh\theta_{\beta_1}:$
\begin{equation}\label{tio2}
\sgn{\lambda_{\beta_1}}\cosh\theta_{\beta_1}=\frac{\pbbe -{\pbae \pbbeae}-{\pbat\pbbeat}}{2\sqrt{\pbae \pbbeae \pbat\pbbeat}}.
\end{equation}
Multiply (\ref{tio2}) with $2\sqrt{\paae \paat}$ and use again that %$p^{b|a}_{\beta_1}=p^{a|b}_{\beta_1},p^{b|a}_{\alpha_1}=p^{a|b}_{\alpha_1}$ and
$\paat=1-\paae$ and ${\bf P}^{a\vert b}$ is double stochastic and
\begin{equation}\label{elva2}
\sgn{\lambda_{\beta_1}}2\sqrt{\paae \paat}\cosh\theta_{\beta_1}=\frac{\paae-1+\pabe+\pbaebe-2\pbaebe \paae}{\sqrt{\pbaebe \pbbeat}} .
\end{equation}
We will show that
$\pm\cosh\left(\gamma_{\alpha_2}-\gamma_{\alpha_1}\right)=\sgn{\lambda_{\beta_1}}\cosh\theta_{\beta_1}$
or equivalently, we show that
 \begin{equation}
\sgn{\lambda_{\beta_1}}2\sqrt{\paae \paat}\cosh\left(\gamma_{\alpha_2}-\gamma_{\alpha_1}\right)=2\sgn{\lambda_{\beta_1}}\sqrt{\paae\paat}\cosh\theta_{\beta_1}.
\end{equation}
We multiply $\pm\sqrt{\paae
\paat}\cosh\left(\gamma_{\alpha_2}-\gamma_{\alpha_1}\right)$ by
$2\sqrt{\paaebe(1-\paaebe)}$ in the left-hand side of (\ref{nia}).
We get $LHS=\pm 2\sqrt{\paaebe(1-\paaebe)}$ $ \sqrt{\paae \paat} $
$\cosh\left(\gamma_{\alpha_2}- \gamma_{\alpha_1}\right)$ and replace
$\sgn{\lambda_{\alpha_1}}\cosh \theta_{\alpha_1}$ by $
\frac{\paae-{\pabe \paaebe}-{(1-\pabe)(1-\paaebe)}}{2\sqrt{\pabe
\paaebe \pabt\paaebt}}$ in the right-hand side
\begin{eqnarray}\label{yes}
LHS %2\sqrt{\paae \paaebe \paat\paaebt}\cos\left(\gamma_{\alpha_2}-\gamma_{\alpha_1}\right)
&=&2\left(2\pabe-1\right)\paaebe(1-\paaebe)\\\nonumber
&+&\left(\paae-{\pabe \paaebe}-{(1-\pabe)(1-\paaebe)}\right)\left(1-2\paaebe\right)\\\nonumber
\end{eqnarray}
We calculate the last term:
\begin{eqnarray}\label{yes2}
&&\left(\paae-1+\pabe+\paaebe-2\pabe\paaebe\right)
\left(1-2\paaebe\right)\\\nonumber
&=&\left(\paae-1+\pabe+\paaebe-2\pabe\paaebe\right)\\\nonumber
&-&2\paaebe\left(\paae-1+\pabe+\paaebe-2\pabe\paaebe\right)\\\nonumber
&=&\left(\paae-1+\pabe+\paaebe\right)-2\pabe\paaebe-2\paaebe \paae\\\nonumber
&-&2\paaebe\left(-1+\pabe+\paaebe-2\pabe\paaebe\right).
\end{eqnarray}
Moreover,
\begin{eqnarray}\label{yes3}
LHS&=&2\left(2\pabe-1\right)\paaebe(1-\paaebe)\\\nonumber
&+&\left(\paae-1+\pabe+\paaebe-2\paaebe \paae\right)\\\nonumber
&-&2\paaebe\left(-1+\pabe+\paaebe-2\pabe\paaebe\right)-2\pabe\paaebe\\\nonumber
&=&2\paaebe\left(-1+2\pabe+\paaebe-2\pabe\paaebe\right)\\\nonumber
&+&\left(\paae-1+\pabe+\paaebe-2\paaebe \paae\right)\\\nonumber
&-&2\paaebe\left(-1+2\pabe+\paaebe-2\pabe\paaebe\right)\\\nonumber
&=&\paae-1+\pabe+\paaebe-2\paaebe \paae.
\end{eqnarray}

Equations (\ref{elva2}) and (\ref{yes3}) imply that
\begin{align}\nonumber
\pm\frac{\paae-1+\pabe+\pbaebe-2\pbaebe\paae }{\sqrt{\pbaebe \pbbeat}}&=\pm\frac{\paae-1+\pabe+\paaebe-2\paaebe \paae}{\sqrt{\paaebe \pabeat}}\\
&\Leftrightarrow\\\nonumber
\pbaebe&=\paaebe.
\end{align}
Therefore we conclude that
$\pm\cosh\left(\gamma_{\alpha_2}-\gamma_{\alpha_1}\right)=
\sgn{\lambda_{\beta_1}}\cosh\theta_{\beta_1}$ iff
$\mathbf{P}^{b|a}=\mathbf{P}^{a|b}$.
Let
\begin{equation}\label{UAB}
U_{b \vert a}^{a \vert b}=\left(
\begin{array}{cc}
 \sqrt{\pbbeae} & \sqrt{\pbbeat} \\
 \sqrt{\pbbtae } & -\sqrt{\pbbtat}
\end{array}
\right).
\end{equation}
We now show that this vector is equivalent to the vector $\psi^{a
\vert b}$ given by (\ref{EXYA}).
\begin{eqnarray} \label{UPSIi}
U_{b \vert a}^{a \vert b} \psi^{b \vert a}
&=& \sqrt{p^a_{\alpha_1}} e^{a\vert b}_{\alpha_1}+ \sgn{\lambda_{\beta_1}}e^{j \theta_{\beta_1}} \sqrt{p^a_{\alpha_2}} e^{a \vert b}_{\alpha_2} \\\nonumber
&=& \sqrt{p^a_{\alpha_1}} e^{a\vert b}_{\alpha_1} +\sgn{\lambda_{\beta_1}}e^{j (\gamma_{\alpha_2}-\gamma_{\alpha_1})} \sqrt{p^a_{\alpha_2}} e^{a \vert b}_{\alpha_2}
\end{eqnarray}
We use the fact that $\psi^{a \vert
b}_{\alpha_i}=\pm\sqrt{p^a_{\alpha_i}}e^{j
\gamma_{\alpha_i}},i\in\{1,2\}$ into (\ref{EXYA})
\begin{eqnarray}
\label{UPSIii}
\psi^{a \vert b}
&=& \pm\sqrt{p^a_{\alpha_1}}e^{j \gamma_{\alpha_1}} e^{a\vert b}_{\alpha_1} \pm \sqrt{p^a_{\alpha_2}} e^{j \gamma_{\alpha_2}}e^{a \vert b}_{\alpha_2} \\\nonumber
&=&\pm e^{i \gamma_{\alpha_1}}U_{b \vert a}^{a \vert b} \psi^{b \vert a}
\end{eqnarray}
\end{proof}
Thus the hyperbolic amplitudes $\psi^{a \vert b}$ and $U_{b \vert
a}^{a \vert b} \psi^{b \vert a}$ differ only by the multiplicative
factor $\pm e^{j \gamma_{\alpha_1}}.$ Hence, they belong to the same
equivalent class of vectors on the unit sphere. Thus they are two
representatives of the same quantum state $\Psi^{b \vert a}.$

\section{Appendix: Hyperbolic algebra and hyperbolic Hilbert space}
\subsection{Hyperbolic algebra}

An element $z$ belongs to the hyperbolic algebra ${\bf G}$ iff it
has following form:
\[z=x+j y,\quad x,y \in \mathbf{R}\]
where $j^2=1$,\; $z_1+z_2=x_1+x_2+j(y_1+y_2)$ and $z_1 z_2=x_1
x_2+y_1+y_2+j(y_1 x_2+y_2 x_1)$. The hyperbolic conjugation is
defined as $\bar{z}=x-j y.$  We define the "square of the absolute
value" as
\[|z|^2=z \bar{z}=x^2-y^2,\] $|z|^2\in{\bf G}.$ In fact,
$|z|^2\in\mathbf{R}$. But $|z|$ is not well defined for $z$ such
that $|z|^2<0$. Therefore set
\[\mathbf{G}_+=\{z\in\mathbf:|z|^2\geq0\}.\] and
\[\mathbf{G}^*_+=\{z\in\mathbf:|z|^2>0\}.\]
We define the argument $\arg{z}$ of $z\in \mathbf{G^*_+}$ as
\[\arg{z}=\text{arctanh}{\frac{y}{x}}=\frac{1}{2}\ln\frac{x+y}{x-y}.\]
Notice that $x\neq0,\;x-y\neq0$ and  $\frac{x+y}{x-y}>0$,\;since $z
\in \mathbf{G}^*_+$.

\subsection{Hyperbolic Hilbert space}

A hyperbolic Hilbert space $H$ is a $\mathbf{G}$-linear inner
product space. Let $x,y,z\in H$ and $a,b\in \mathbf{G}$, then
consider the inner product as a map from $H\times H\rightarrow
\mathbf{G}$ having the following properties:

 \medskip
(1) Conjugate symmetry: $\left\langle x,y\right\rangle$ is the
conjugate to $\left\langle y,x\right\rangle$
\[\left\langle x,y\right\rangle=\overline{\left\langle y,x \right\rangle}\]

(2) Linearity with respect to  the first argument:
\[\left\langle a x +b z,y\right\rangle=a\left\langle x,y\right\rangle+b\left\langle z,y\right\rangle\]

(3) Nondegenerate:
  \[\left\langle x,y\right\rangle=0\]
for all $y\in H$ iff $x=0$ \\
In general, the norm $\left\|\psi \right\|=\sqrt{\left\langle \psi,
\psi\right\rangle}$ is not well defined. But we will only need  the
square of the norm $\left\|\psi \right\|^2=\left\langle \psi, \psi
\right\rangle$.

\subsection{Violation of Born's rule} \label{VBR}
Let us give a counterexample to illustrate the violation of Born's rule, if the transition probabilities matrix ${\bf P}^{b \vert a}$ is not doubly stochastic. We have that \begin{equation}
\label{CE1} \psi^{b\vert a}_\beta \;=\; \sqrt{p^a_{\alpha_1}p^{b\vert a}_{\beta \alpha_1}}
\pm  e^{j \theta_\beta} \sqrt{p^a_{\alpha_2} p^{b\vert a}_{\beta \alpha_2}}\,, \quad
\beta \in X_b\,\end{equation} and
\begin{equation}
\label{CE2}
\psi^{b \vert a}  =
\psi^{b\vert a}_{\beta_1} e^{b\vert a}_{\beta_1} + \psi^{b\vert a}_{\beta_2} e^{b\vert a}_{\beta_2}.
\end{equation}
This will match Born's rule,
\begin{equation}
\label{CE3}
p^b_{\beta}=\vert \langle \psi^{b\vert a}, e^{b \vert a}_\beta
\rangle\vert^2 \;, \beta \in  X_b.
\end{equation}
Moreover $p^b_{\beta_1}+p^b_{\beta_2}=1$ and by \eqref{PBbeta}(3.1),
\begin{align}
\label{CE4}
1=p^b_{\beta_1}+p^b_{\beta_2}&=p^a_{\alpha_1}(p^{b\vert a}_{\beta_1 \alpha_1}+p^{b\vert a}_{\beta_2 \alpha_1})+p^a_{\alpha_2}(p^{b\vert a}_{\beta_1 \alpha_2}+p^{b\vert a}_{\beta_2 \alpha_2})\\\nonumber &+2 \sqrt{p^a_{\alpha_1}p^a_{\alpha_2}}(\lambda_1\sqrt{p^{b\vert a}_{\beta_1 \alpha_1}p^{b\vert a}_{\beta_1 \alpha_2}}+\lambda_2\sqrt{p^{b\vert a}_{\beta_2 \alpha_1}p^{b\vert a}_{\beta_2 \alpha_2}}).
\end{align}
Let us select the transition probabilities matrix ${\bf P}^{b \vert a}$ not to be doubly stochastic, take $p^{b\vert a}_{\beta_1 \alpha_1}=p^{b\vert a}_{\beta_1 \alpha_2}=p$ and $p^{b\vert a}_{\beta_2 \alpha_1}=p^{b\vert a}_{\beta_2 \alpha_2}=q$ where $p+q=1,\; p\neq q,\; p,q>0.$
Then \eqref{CE4} becomes
\begin{equation}
\label{CE5}
1=p^a_{\alpha_1}+p^a_{\alpha_2}+2 \sqrt{p^a_{\alpha_1}p^a_{\alpha_2}}(\lambda_1 p+\lambda_2 q)\Leftrightarrow
\lambda_1=-\frac{q}{p}\lambda_2
\end{equation}
Then \eqref{PSI} (3.11) will be \begin{equation}
\label{PSICE}
\psi^{b\vert a}= \sqrt{p^a_{\alpha_1}} e^{b\vert a}_{\alpha_1} \pm
e^{j \theta_{\beta_1}} \sqrt{p^a_{\alpha_2}} e^{b\vert a}_{\alpha_2}
\end{equation}
where
\begin{equation}
 e^{b\vert a}_{\alpha_1} \;=\;\left( \begin{array}{l} \sqrt{p^{b\vert a}_{\beta_1 \alpha_1}} \\
\sqrt{p^{b\vert a}_{\beta_2 \alpha_1}}
\end{array}
\right ),\quad e^{b\vert a}_{\alpha_2} \;=\;\left( \begin{array}{l}  \; \sqrt{p^{b\vert a}_{\beta_1 \alpha_2}} \\
- \frac{q}{p}\sqrt{p^{b\vert a}_{\beta_2 \alpha_2}}
\end{array}
\right )\,.
\end{equation}
Thus $p^b_{\beta}=\vert \langle e^{b \vert a}_{\alpha_1}
\rangle, e^{b \vert a}_{\alpha_2}
\rangle\vert^2=p-\frac{q^2}{p} \;,$ where $e^{b \vert a}_{\alpha_1}$ and $e^{b \vert a}_{\alpha_2}$, are orthogonal, i.e. $\vert \langle e^{b \vert a}_{\alpha_1}
\rangle, e^{b \vert a}_{\alpha_2}
\rangle\vert^2=0$.
This shows the violation of Born's rule by contradiction, since $p-\frac{q^2}{p}=0\Leftrightarrow p=\pm q$ . 
\medskip
\subsection*{Acknowledgment}

I am grateful to my supervisor Professor Andrei Khrennikov for
discussions  and introducing me into this field of research. I am
also very thankful to Guillaume Adenier for discussions on
foundations  of quantum mechanics.

\end{document}